\documentclass[useAMS]{mn2e}

\usepackage{ulem} 
\usepackage{graphicx}
\usepackage{amsmath}
\usepackage{ulem}
\usepackage{rotating}
\usepackage{amssymb}

\def\lapprox{\hbox{\lower .8ex\hbox{$\,\buildrel < \over\sim\,$}}}
\def\gapprox{\hbox{\lower .8ex\hbox{$\,\buildrel > \over\sim\,$}}}

\title[\text{HST} on NGC\,6752. I. A new dwarf Galaxy in background]{
The \textit{HST} Large Programme on NGC\,6752. I. Serendipitous discovery of a dwarf Galaxy in background\thanks{
Based on observations with the NASA/ESA {\it Hubble
Space Telescope}, obtained at the Space Telescope Science Institute,
which is operated by AURA, Inc., under NASA contract NAS 5-26555.
}
}
\author[L.\,R.\,Bedin et al.]{
  L.\,R.\,Bedin$^{1}$\thanks{E-mail: luigi.bedin@oapd.inaf.it}, M.\,Salaris$^{2}$, R.\,M.\,Rich$^{3}$,
  H.\,Richer$^{4}$, J.\,Anderson$^{5}$, D.\,Bettoni$^{1}$, 
\newauthor
  D.\,Nardiello$^{6}$, A.\,P.\,Milone$^{6}$, A.\,F.\,Marino$^{6}$, M.\,Libralato$^{5}$, A.\,Bellini$^{5}$, A.\,Dieball$^{7}$,  
\newauthor
  P.\,Bergeron$^{8}$, A.\,J.\,Burgasser$^{9}$  and D.\,Apai$^{10,11}$\\
$^{1}$INAF-Osservatorio Astronomico di Padova, Vicolo dell'Osservatorio 5, I-35122 Padova, Italy\\
$^{2}$Astrophysics Research Institute, Liverpool John Moores University,146 Brownlow Hill, Liverpool L3 5RF, UK\\
$^{3}$Department of Physics and Astronomy, UCLA, 430 Portola Plaza, Box 951547, Los Angeles, CA 90095-1547, USA\\
$^{4}$Department of Physics and Astronomy, University of British Columbia, Vancouver, BC, V6T 1Z1, Canada\\
$^{5}$Space Telescope Science Institute, 3800 San Martin Drive, Baltimore, MD 21218, USA\\ 
$^{6}$Dipartimento di Fisica e Astronomia ‘Galileo Galilei’, Università di Padova, Vicolo dell’Osservatorio 3, Padova I-35122, Italy\\
$^{7}$Argelander Institut f\"ur Astronomie, Helmholtz Institut f\"ur Strahlen-und Kernphysik, University of Bonn, Germany\\
$^{8}$D\'epartement de Physique, Universit\'e de Montr\'eal, C.P.\,6128, Succ.\,Centre-Ville, Montr\'eal, QC\,H3C\,3J7, Canada\\
$^{9}$Center for Astrophysics and Space Science, University of California San Diego, La Jolla, CA 92093, USA\\
$^{10}$Department of Astronomy and Steward Observatory, The University of Arizona, 933 N. Cherry Avenue, Tucson, AZ 85721, USA\\
$^{11}$Lunar and Planetary Laboratory, The University of Arizona, 1640 E. University Blvd., Tucson, AZ 85721, USA
}

\begin{document} 
\date{Accepted 2019 January 8. Received 2019 January 8; in original form 2018 November 28}

\pagerange{\pageref{firstpage}--\pageref{lastpage}} \pubyear{201X}

\maketitle
 
\label{firstpage}

\begin{abstract}
As part of a large \textit{Hubble Space Telescope} investigation
aiming at reaching the faintest stars in the Galactic globular cluster
NGC\,6752, an ACS/WFC field was the subject of deep optical
observations reaching magnitudes as faint as $V\sim30$.  In this field
we report the discovery of \textit{Bedin\,I},
a dwarf spheroidal galaxy too faint and too close to the core of
NGC\,6752 for detection in earlier surveys.
As it is of broad interest to complete the census of galaxies in the
local Universe, in this Letter we provide the position of this new
object along with preliminary assessments of its main parameters.
Assuming the same reddening as for NGC\,6752, we estimate a distance
modulus of $(m-M)_0=29.70\pm0.13$ from the observed red giant branch,
i.e., 8.7$^{+0.5}_{-0.7}$\,Mpc, and size of $\sim$840$\times$340\,pc,
about 1/5 the size of the LMC. A comparison of the observed
colour-magnitude diagram with synthetic counterparts that account for
the galaxy distance modulus, reddening and photometric errors,
suggests the presence of an old ($\sim$13~Gyr) and metal poor
([Fe/H]$\sim$$-$1.3) population.  This object is most likely a
re\-la\-tively isolated satellite dwarf spheroidal galaxy of the
nearby great spiral NGC\,6744, or potentially the most distant
isolated dwarf spheroidal known with a secure distance.
\end{abstract}


%
\section{Introduction}
\label{introduction}
%
%

New technologies and increasingly deep and wide surveys have resulted
in the recent discovery of numerous dwarf spheroidal galaxies,
principally in the vicinity of the Milky Way or M\,31.
Only a handful of dwarf spheroidal galaxies with well established
distances appear to be truly isolated.
Cetus lies at 775 kpc from the Milky Way and 680 kpc from M\,31
(Whiting, Hau, \& Irwin 1999; Lewis et al.  2007), Tucana is roughly
900 kpc from the Milky Way and 1350 kpc from M\,31 (Lavery \& Mighell
1992; Saviane et al.\ 1996), and KKR\,25 at about 2\,Mpc from both
spirals is one of the most isolated (Makarov et al.\ 2012).
The smallest of these stellar systems can exhibit either complex star
formation histories, or more closely resemble the purely old,
$\sim$10\,Gyr po\-pu\-lations of the extended globular clusters in
M\,31 (Huxor et al.\ 2005; Mackey et al.\ 2006).  Although a number of
possible additional cases are cited in Martinez-Delgado et
al.\ (2018), the candidates usually lack secure distances, as the tip
of the red giant branch (TRGB) becomes (under most circumstances) very
difficult to measure at distances beyond 4 Mpc, even with
\textit{Hubble Space Telescope} \textit{(HST)} imagery.  Any complex
history of star formation makes the TRGB even harder to discern.

Here we report the discovery of one of the most distant, relatively
isolated dwarf spheroidal galaxy with a secure TRGB distance.  We have
discovered this object serendipitously with extraordinarily deep
\textit{HST} images obtained for the purpose of investigating the
white dwarf cooling sequence of the globular cluster NGC\,6752.

\begin{figure*}
\begin{center}
\includegraphics[width=168mm]{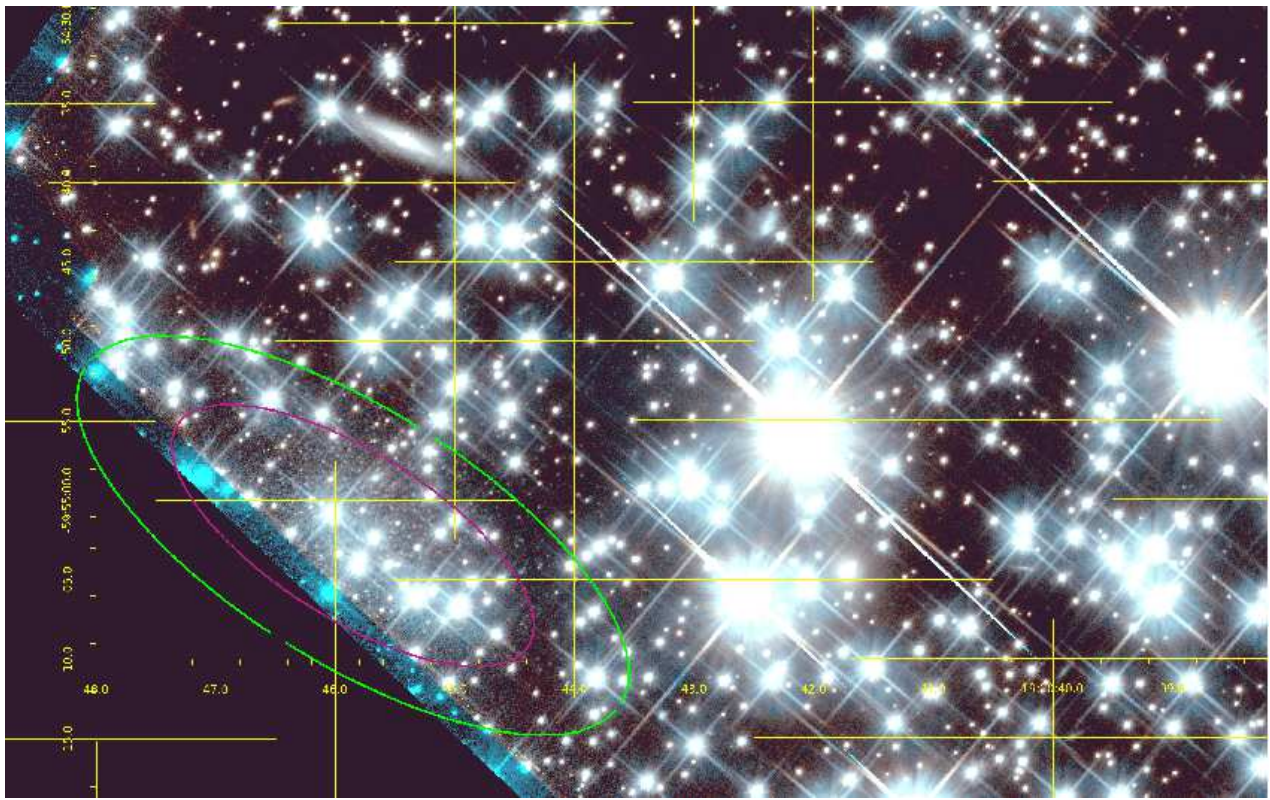}
\caption{
  A $80^{\prime\prime}\times50^{\prime\prime}$ portion of the ACS/WFC
  field containing Bedin\,I. North is up, East to the left (ICRS
  coordinates grid in yellow). Unfortunately, Bedin\,I is very close
  to the border of the field of view (fov), where the large dithers
  gave us an overall incomplete (particularly in the red filter) and
  shallower view with respect to the centre of the fov.  This portion
  of the field contains what is available to us of the new dwarf
  spheroidal, and extends West to show an adjacent region for
  comparison. The green ellipse marks the limit to where Bedin\,I
  seems to extend, while the one in magenta denotes the fitted
  half-light radius (see text).
\label{stack}
}
\end{center}
\end{figure*}
%
%

%
\section{Observations} 
%

All images for this study were collected with the \textit{Wide Field
  Channel} (WFC) of the \textit{Advanced Camera for Surveys} (ACS) at
the focus of the \textit{HST} under program GO-15096 (PI: Bedin).
Unfortunately, five out of the planned 40 orbits failed because of
poor guide star acquisition and will be reobserved at a later time.
Usable data were collected between September 7 and 18, 2018, and
consist of deep exposures of 1270\,s each, 19 in the F814W filter, and
56 in F606W.
Note that this is an astrometric multi-cycle programme, and a second
epoch (also of 40 orbits) has already been approved (GO-15491) and
scheduled for late 2019.  Proper motions will eventually provide a
near perfect field-object decontamination both for NGC\,6752 members
in the foreground, and for the members of the newly and
serendipitously discovered stellar system in the background, to which
we refer hereafter in this work as \textit{Bedin\,I}.
This object is worthy of early and separate publication as it is
likely of general interest to the community as an example of a
relatively isolated extragalactic system.

For completeness, Paper\,III of this series is focused on the white
dwarfs of NGC\,6752 in this very same ACS/WFC field, while Paper\,II
deals with multiple stellar populations detected within NGC\,6752 in
our parallel observations with the \textit {Infra-Red channel} (IR) of
the \textit{Wide Field Camera 3} (WFC3).
%

%
\section{Data Reduction and Analysis} 
%

All images were pre-processed with the pixel-based correction for
imperfections in the charge transfer efficiency (CTE) with the method
described in Anderson \& Bedin (2010).
Photometry and relative positions were obtained with the software
tools described by Anderson et al.\ (2008). In addition to solving for
positions and fluxes, important diagnostic parameters were also
computed, such as the image-shape parameter, which quantifies the
fraction of light that a source has outside the predicted point-spread
function (PSF).  This is useful for eliminating the faint unresolved
galaxies that tend to plague studies of faint point sources.

The astrometry was registered to International Celestial Reference
System (ICRS) using sources in common with Gaia\,DR2 (Gaia
collaboration 2018) with tabulated proper motions transformed to the
epoch 2018.689826 of \textit{HST} data, following the procedures in
Bedin \& Fontanive (2018).

The photometry was calibrated on the ACS/WFC Vega-mag system following
the procedures given in Bedin et al.\ (2005) using encircled energy
and zero points available at
STScI.\footnote{\texttt{http://www.stsci.edu/hst/acs/analysis/zeropoints}}
For these calibrated magnitudes we will use the symbols
$m_{\rm  F606W}$ and $m_{\rm F814W}$.

The software also corrects for distortion in all the images and
transforms their coordinates to a common reference frame after removal
of cosmic rays and most of the artifacts.
It then combines them to produce stacked images that, at any location,
give sigma-clipped values of the individual values from pixels of all
images at that location.

In Fig.\,\ref{stack} we show a portion of a pseudo-trichromatic stack
containing Bedin\,I.  [Note that we adopted the F814W images for the
  red channel, the F606W ones for the blue channel, and computed a
  wavelength weighted (blue/red$\sim$3) average of the two for the
  green channel.]
In these images the dwarf spheroidal clearly appears.  Unfortunately
the system is close to the border of the field, where the depth of the
coverage is lowered by the large dithers strategy employed for this
programme, and its view is incomplete (particularly in F814W).

%
We attempted to fit an ellipse to derive the luminosity profile
(Jedrzejewski et al.\ 1987), however there are several much brighter
foreground stars superimposed on the faint galaxy.  To be able to fit
the underlying galaxy, we masked all contaminating sources by hand as
well as all the visible diffraction spikes.  We used the images in the
F606W filter, as they cover a larger area and reach deeper magnitude
than those in F814W.  A single Sersic function (Sersic 1963, see also
Graham et al.\ 2005) was used as model, allowing all parameters to
vary.  The final fit gives parameters for Bedin\,I listed in
Table\,\ref{TAB}. However, these values should be treated as
indicative, given the level of contamination and the incomplete view
of the system. Figure\,\ref{stack} show the fitted ellipse at an
extension where the profile drops to zero (in green), and at the
half-luminosity radius (in magenta).
%

\begin{figure}
\begin{center}
\includegraphics[width=88mm]{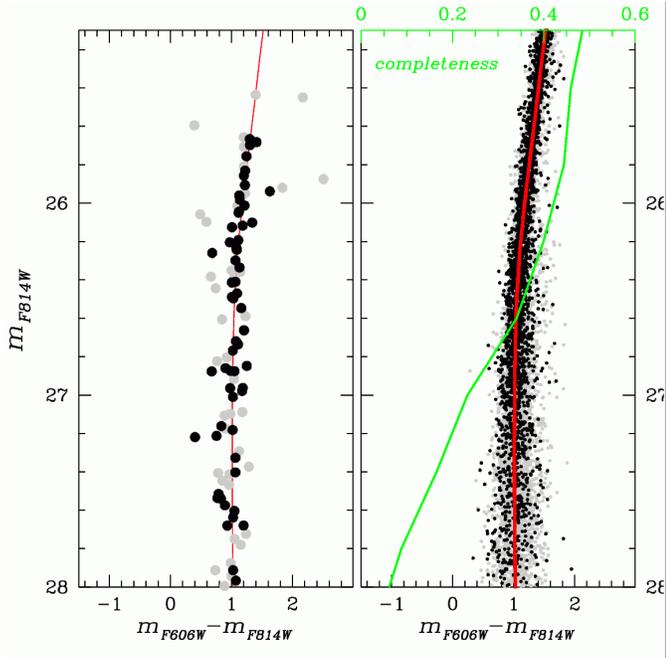}
\caption{
  \textit{(Left panel:)} CMD of all stars in the region of Bedin\,I
  (grey) and for just those ones closest to its centre (in black, see
  text). The fiducial line defined by-eye and used to generate
  artificial stars, is displayed in red.
  \textit{(Right panel:)} Same CMD but for the artificial stars in the
  same regions (using same colour code) added along the fiducial line
  (in red). We also show the derived completeness function (in green),
  whose values can be read on the top axis.
\label{RGBt}
}
\end{center}
\end{figure}

Given the uncertainties of this fit, and the physical linear edge of
the fov, we define the dwarf-galaxy sample within boxes rather than
ellipses.
In the left panel of Fig.\,\ref{RGBt} we show the colour-magnitude
diagram (CMD) for all measured stars within a box of
75$^{\prime\prime}\times 110^{\prime\prime}$ containing Bedin\,I (grey
points), and the closest one to the centre of the system, within a
15$^{\prime\prime}\times 30^{\prime\prime}$ box (black points).
This CMD shows clearly a red giant branch (RGB) that appears to end
around $m_{\rm F814W}$=25.6-25.8.
We then defined by-eye a fiducial line (shown in red) representative
of the RGB, extending beyond the apparent TRGB.

Next, we performed artificial-star tests (ASTs) within the same box
containing Bedin\,I; these ASTs were performed using the procedures
described by Anderson et al.\ (2008). Artificial stars were added in a
magnitude range $m_{\rm F814W}$=20-29 with a flat distribution, with
colours that placed them on the fiducial RGB, to determine
completeness and to estimate errors; the resulting CMD for ASTs is shown
in the right panel of Fig.\,\ref{RGBt} along with the derived
completeness (green line).
[Note the rather low completeness --$\sim$45\%-- at the TRGB level,
  because a large fraction of the area is disturbed by the halos of
  saturated stars, see Fig.\,\ref{stack}.]

%
\begin{table}
\caption{Properties of \textit{Bedin\,I} dwarf spheroidal galaxy.}
\center
\begin{tabular}{lc}
\hline
 & \\
$\alpha_{\rm ICRS}$      &   19$^{\rm d}$:10$^{\rm m}$:45$^{\rm s}$.85  \\
$\delta_{\rm ICRS}$      &$-$59$^{\circ}$:55$^{\prime}$:02$^{\prime\prime}$.25 \\
$\ell$ &  336\,.\!\!$^{\circ}$56365 \\
$b$    &$-$25\,.\!\!$^{\circ}$60333 \\
 & \\
$r_{e}$ (\textit{effective} half-light radius) & $\sim$8.3$^{\prime\prime}$ ($\sim$350\,pc) \\
$\epsilon = \frac{a-b}{a}$    & 0.6  \\
$a_{\rm max}$-axis& $\sim$20$^{\prime\prime}$ ($\sim$840\,pc) \\
$b_{\rm max}$-axis& $\sim$8$^{\prime\prime}$ ($\sim$340\,pc)\\
P.A.    & 58$^{\circ}$  \\
 & \\
$E(m_{\rm F606W}-m_{\rm F814W})$ & 0.04 \\
$(m-M)_0$ & 29.70$\pm$0.13 \\
$d$ & 8.7$^{+0.5}_{-0.7}$\,Mpc\\
$m_{\rm F606W}$   &    19.94 \\
$M_{\rm F606W}$   &  $-$9.76 \\
$\mu_{\rm F606W}$   &  26.78\,mag/arcsec$^2$ \\
 & \\
\hline
\end{tabular}
\label{TAB}
\end{table} 
%

%

%
\section{Distance and metallicity} 
%
%
Based on the data displayed in Fig.~\ref{RGBt} we have made a rough
formal estimate of distance and metallicity of this galaxy.  For the
distance we employed the TRGB absolute magnitude-colour relation in
the ACS/WFC system by Rizzi et al.~(2007):
\begin{equation}
M_{\rm F814W}^{\rm TRGB}=-4.06+0.20[(m_{\rm F606W}-m_{\rm F814W})_{\rm TRGB}-1.23]
\label{TRGBcalib}
\end{equation}
An edge detection algorithm with kernel [$-$1 0 +1] (Sobel filter,
see, e.g., Madore \& Freedman~1995) was employed to determine the TRGB
$m_{\rm F814W}$ apparent magnitude, considering all stars within the
75$^{\prime\prime}\times 110^{\prime\prime}$ box containing Bedin\,I
(black and grey points in Fig.~\ref{RGBt}). There are about 50 stars
(without accounting for the effect of completeness) in the upper
magnitude bin below $m_{\rm F814W}$=25.5, that seems to be
approximately the location of the TRGB. With this sample size,
systematic errors on the location of the TRGB using a Sobel filter are
below 0.1~mag (Madore \& Freedman~1995).

We have determined the $m_{\rm F814W}$ differential luminosity
function (LF) for the RGB stars with $m_{\rm F814W}$ between $\sim$23
and 28~mag, in 0.25~mag bins (this width corresponds to 2-3 times the
1$\sigma$ photometric error estimated from the ASTs for $m_{\rm F814W}<$27.0),
and applied our derived completeness corrections to star counts in each bin.

\begin{figure}
\begin{center}
\includegraphics[width=88mm]{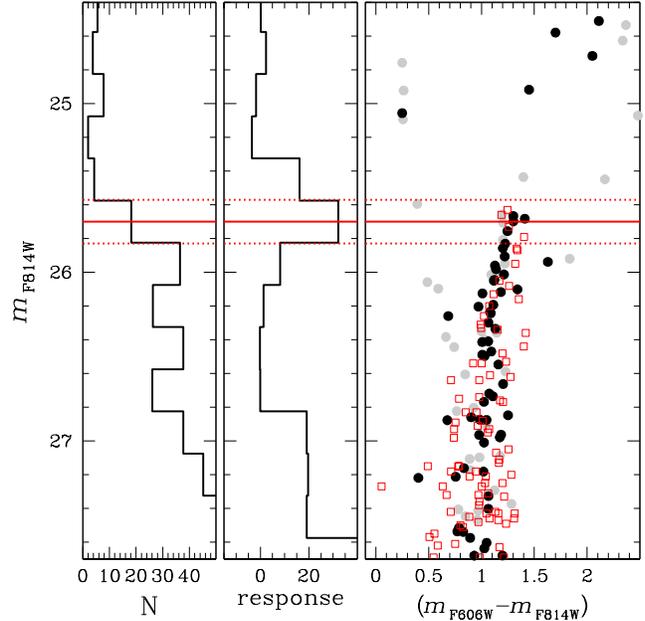}
\caption{
The left and middle panel display, respectively, the observed LF of
the galaxy upper RGB, and the output of the edge detection
algorithm. The position of the TRGB is marked with a red solid line,
while the dotted lines give the uncertainty (see text for details).
The right panel displays the galaxy RGB (symbols as in
Fig.~\ref{RGBt}) and a synthetic CMD (red squares) for a 13\,Gyr old,
[Fe/H]=$-$1.27 population (see text for details).
\label{trgb}
}
\end{center}
\end{figure}

We then applied the Sobel's edge-detection
algorithm to this completeness corrected LF; a sharp spike in the
filter output (see middle panel of Fig.~\ref{trgb}) marks the position
of TRGB, that we take as the midpoint of the bin corresponding to the
spike.  We have repeated several times this procedure by changing the
magnitude of the initial point of the brightest bin by 0.01~mag at a
time, and determined $m_{\rm F814W}^{\rm TRGB}=25.70\pm0.13$.

As for the TRGB colour to employ in Eq.~\ref{TRGBcalib}, an histogram
of the colour distribution (0.1~mag bins) for all stars with $m_{\rm
  F814W}$ between 25.53 and 25.83~mag discloses a clear peak at
$(m_{\rm F606W}-m_{\rm F814W})_{\rm TRGB}=1.22\pm0.05$ that we take as
the galaxy TRGB mean colour.

We assumed a reddening $E(B-V)$=0.04 as for NGC\,6752 (Harris~1996,
updated by December 2010), corresponding to ${\rm A_{F606W}}$=0.11,
${\rm A_{F814W}}$=0.07, following the extinction law in the ACS/WFC
photometric system determined by Bedin et al.~(2005). This gives a
dereddened TRGB colour $(m_{\rm F606W}-m_{\rm F814W})_{\rm
  TRGB}=1.18\pm0.05$.  By using the TRGB absolute magnitude
calibration of Eq.~\ref{TRGBcalib} we obtained a distance modulus
$(m-M)_{0}=29.70\pm0.13$, corresponding to a linear distance of about
$8.7^{+0.5}_{-0.7}$~Mpc.

Figure~\ref{trgb} also displays a qualitative comparison of the galaxy
CMD with a synthetic counterpart\footnote{The number of synthetic
  stars below the model TRGB is the same as the number of observed
  stars below the TRGB as identified by the edge detection
  algorithm.}, obtained employing scaled-solar BaSTI isochrones
(Pietrinferni et al.~2004) for an age of 13\,Gyr and [Fe/H]=$-$1.27.
We have included photometric errors in both $m_{\rm F814W}$ and
$m_{\rm F606W}$ as derived from the ASTs, and shifted the CMD of the
synthetic population by $(m-M)_{0}=29.70$, ${\rm A_{F606W}}$=0.11 and
${\rm A_{F814W}}$=0.07.
The synthetic CMD nicely overlaps with the observational counterpart.
Its colour width is quite comparable with the observed one, and the
theoretical TRGB agrees with the edge detection results.  This
suggests that, within the current photometric errors, the observed
galaxy population is quite homogeneous, similar to typical metal poor
globular clusters.

Notably, the CMD resembles that of a number of M31 dwarf
spheroidal galaxies (Martin et al.\ 2017).  It is unlikely that any
new observations might reach the red clump/horizontal branch, that
could contain interesting information on the system's age.

%
%
%

%
\section{Conclusions}
%

We conclude that this object is most likely an isolated dwarf
spheroidal galaxy, but could be associated with NGC\,6744.
Indeed, NGC\,6744 has a distance of 9.15$\pm$0.40\,Mpc (Tully et
al.\ 2013 catalog gives $(m-M)_0$=29.81$\pm$0.09),
i.e., consistent with the dwarf within uncertainties, and it is at an
angular distance of about 4\,degrees from the dwarf spheroidal (so, at
a minimum distance of $\sim$650\,kpc on the plane of the sky).
The galaxy NGC\,6744 is a Milky-Way-analog barred spiral with
$M_V=-$21.8\footnote{\texttt{ned.ipac.caltech.edu}}, a member of the
Pavo group, and has a low luminosity AGN (da Silva et al.\ 2018).

The size and the estimated ellipticity of Bedin\,I offer the closest 
resemblance to the Tucana dwarf spheroidal galaxy (see e.g.,
Fraternali et al.\ 2009).
KKR\,25 is an almost identical isolated dwarf system, with similar
luminosity ($M_V\sim-$10.93), ellipticity, and color-magnitude diagram
(Makarov et al.\ 2012).
The new system is both too large, and insufficiently compact, to be
classified alongside the extended globular clusters of M\,31 (Huxor et
al. 2005).  At $M_{\rm F606W}\sim -10$, the system would be among the
more luminous M\,31 dwarf spheroidal galaxies.  The half-light
effective radius ($r_e\sim$350\,pc) and ellipticity place it
comfortably among the M\,31 dwarfs spheroidals (Martin et al.\ 2017)
and the Milky Way dwarfs (Collins et al.\ 2014).
The system also appears to have had a quiet star formation history,
with a red giant branch resembling that of the majority of M\,31 dwarf
spheroidal galaxies (Martin et al.\ 2017).  The narrow RGB and
[Fe/H]=$-$1.3 admits little internal production of metals, and our
limited sample of sufficiently well-measured stars show no obvious or
sizable population of AGB or younger stars.
Our photometry is too shallow to reach the horizontal branch, that
would provide additional constraints on both the distance and star
formation history, and such faint magnitudes may be beyond any current
or contemplated facilities.
Considering the isolation of the system relative to any luminous
galaxy, it is curious that the RGB is extremely narrow, implying
little internal production of metals, suggesting a chemical evolution
closer to that of globular cluster than a fully fledged dwarf galaxy.
We conclude that this system likely formed more than 10\,Gyr ago in a
``single'' burst, and likely experienced no additional star formation
since its formation.  The reasonably well-determined distance of this
system makes it one of the best candidates for a relatively isolated
dwarf spheroidal galaxy.

As a final remark, it is interesting to compare Bedin\,I with the
faint dwarf irregular galaxy \textit{LV\,J1157+5638\,sat} discovered
by Makarova et al.\ (2018).  A comparison to Fig.\,5 of Makarova et
al.\ (2018) shows that Bedin\,I is very likely among the least
luminous galaxies known at a distance of more than 4\,Mpc.  Both
systems were discovered serendipitously in deep \textit{HST} imaging.
These two works bode well for the potential of the High Latitude
Survey planned for the Wide Field Infrared Survey Telescope
\textit{(WFIRST)}\footnote{\texttt{https://wfirst.gsfc.nasa.gov/,
    https://www.wfirst-hls-cosmology.org/}} to reveal a statistical
population of similar objects.

\section*{Acknowledgments}
We are grateful to an anonymous referee for his/her prompt review of
our work.
This work is based on observations with the NASA/ESA Hubble Space
Telescope, obtained at the Space Telescope Science Institute, which is
operated by AURA, Inc., under NASA contract NAS 5-26555.
J.A., R.M.R., A.B, M.L, and acknowledge support from \textit{HST}-GO-15096.
%
A.P.M. acknowledge funding from the European Research Council (ERC)
under the European Union's Horizon 2020 research innovation programme
(Grant Agreement ERC-StG 2016, No 716082 'GALFOR'.
APM acknowledges support from MIUR through the the FARE project
R164RM93XW ‘SEMPLICE’.
AFM has received funding from the European Union’s Horizon 2020
research and innovation programme under the Marie Sk{\l}odowska-Curie
Grant Agreement No.\,[797100].

%


\label{lastpage}


\end{document}